# A monitoring tool for a GRID operation center

S. Andreozzi
*INFN-CNAF, Bologna, I-40126, Italy*

S. Fantinel
*INFN-LNL, Legnaro, I-35020, Italy*

D. Rebatto, L. Vaccarossa
*INFN, Milano, I-20133, Italy*

G. Tortone
*INFN, Napoli, I-80126, Italy*

WorldGRID is an intercontinental testbed spanning Europe and the US integrating architecturally different Grid implementations based on the Globus toolkit. The WorldGRID testbed has been successfully demonstrated during the WorldGRID demos at SuperComputing 2002 (Baltimore) and IST2002 (Copenhagen) where real HEP application jobs were transparently submitted from US and Europe using "native" mechanisms and run where resources were available, independently of their location. To monitor the behavior and performance of such testbed and spot problems as soon as they arise, DataTAG has developed the EDT-Monitor tool based on the Nagios package that allows for Virtual Organization centric views of the Grid through dynamic geographical maps. The tool has been used to spot several problems during the WorldGRID operations, such as malfunctioning Resource Brokers or Information Servers, sites not correctly configured, job dispatching problems, etc. In this paper we give an overview of the package, its features and scalability solutions and we report on the experience acquired and the benefit that a GRID operation center would gain from such a tool.

## 1. INTRODUCTION

The ability to monitor and manage distributed computing components is critical for enabling high-performance distributed computing.

Monitoring data is needed:
- to determine the source of performance problems and to tune the system for better performance
- to enable fault detection and recovery mechanism
- as inputs for a performance prediction service

As distributed systems such as Computational Grids become bigger, more complex, and more widely distributed, it becomes important that this monitoring and management be automated.

Monitoring information is collected in Grid "snapshots" that should include specific information for each kind of resource:
- network devices (routers, switches, etc.): CPU load, memory, in bandwith, out bandwith, …
- computing element: CPU load, memory, number of users, number of processes, …
- storage element: free space, number of I/O interrupts, …
- network services: DNS, rsh/ssh, telnet, NFS, …
- Grid services: gatekeeper, job-manager, replica catalog, GSIftp …

## 2. THE WORLDGRID TESTBED

The WorldGRID testbed is a first successful example of Grid interoperability across EU and US domains; it is a basic collaboration between European and US Grid projects in order to demonstrate the interoperability between Grid domains for applications submitted by users from different Virtual Organizations. The Grid projects involved in this activity are: iVDGL[1], DataGRID[2], CrossGrid[3] and DataTAG[4].

## 3. WORLDGRID TESTBED MONITORING

To monitor the behavior and performance of such testbed and spot problems as soon as they arise, the Work Package 4 of DataTAG project has developed the EDT-Monitor tool based on the Nagios[5] package that allows for Virtual Organization centric views of the Grid through dynamic geographical maps.

The extremely flexible interface of Nagios, permits to produce powerful plugins that can add features to the Nagios core.

### 3.1. Nagios features

Nagios is an OpenSource host and service monitor designed to inform about network problems. It has been designed to run under the Linux operating system, but works fine under most *NIX variants as well. The monitoring daemon runs intermittent checks on hosts and services you specify using external "plugins" which return status information to Nagios. When problems are encountered, the daemon can send notifications out to administrative contacts in a variety of different ways (email, instant message, SMS, etc.). Current status information, historical logs, and reports can all be accessed via a web browser.

Main features included:
- monitoring of network services
- monitoring of host resources
- simple plugin design that allows users to easily develop their own host and service checks
- ability to define network host hierarchy, allowing detection of and distinction between hosts that are down and those that are unreachable

- contact notifications when service or host problems occur and get resolved (via email, pager, or other user-defined method)
- optional escalation of host and service notifications to different contact groups
- ability to define event handlers to be run during service or host events for proactive problem resolution
- support for implementing redundant and distributed monitoring servers
- external command interface that allows on-the-fly modifications which must be made to the monitoring and notification behavior through the use of event handlers, the web interface, and third-party applications
- retention of host and service status across program restarts
- scheduled downtime for suppressing host and service notifications during periods of planned outages
- ability to acknowlege problems via the web interface
- web interface for viewing current network status, notifications and problems history, logs file, etc.
- simple authorization scheme that allows you restrict what users can see and do from the web interface

### 3.2. EDT-monitor features

We implemented plugins that deliver information in a more suitable graphic manner.

One of these permits to represent sites and corresponding metrics over a world map. Dots with different colors are displayed to display a metric's site. For example if a site has a heavy CPU, the corresponding dot for that site will be red on the map, otherwise it will be green (Figure 1). The plugins that have been developed are related to the following metrics: CPU, disk, mem, processes for single Grid element and a MDS plugin that is able to retrieve information from a CE/SE GRIS.

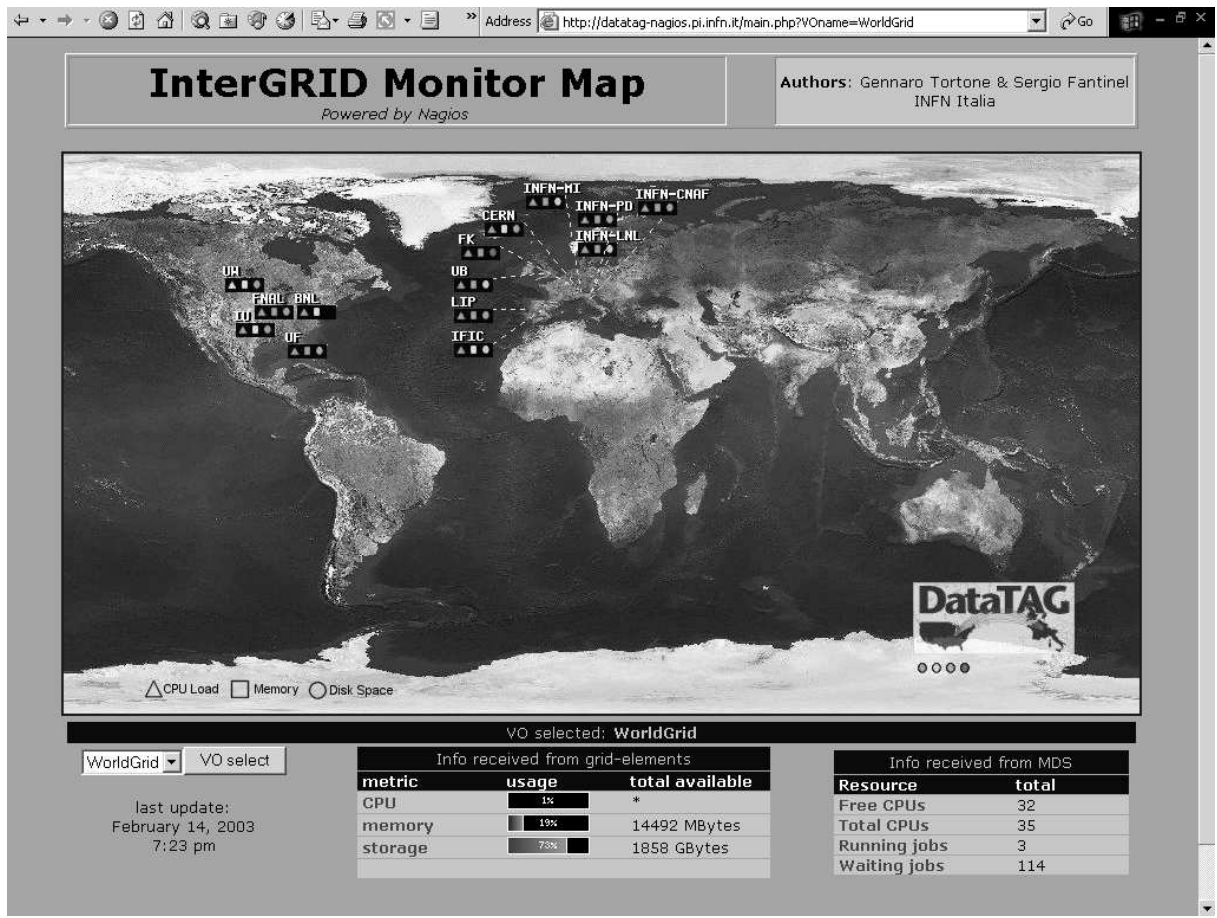

Figure 1: WorldGRID monitoring map

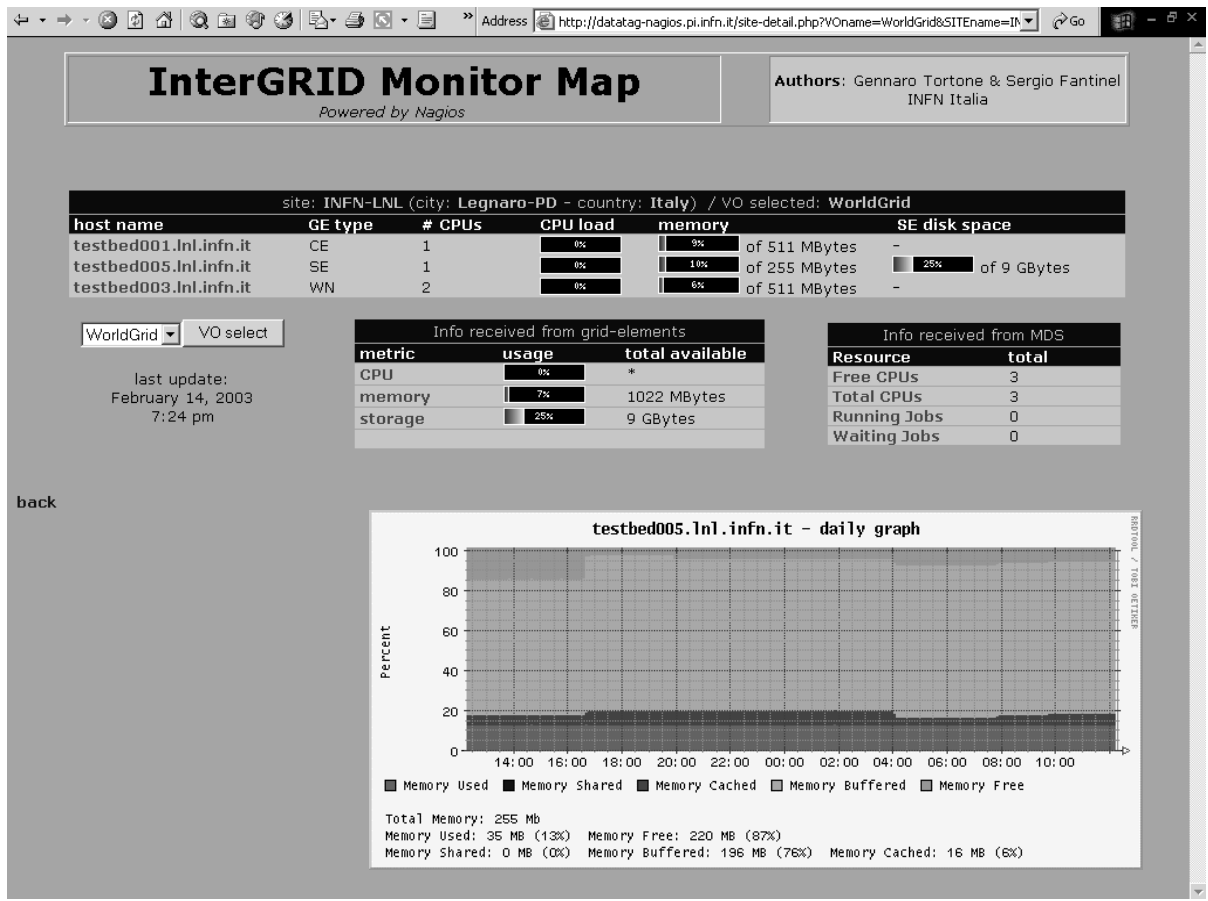

Figure 2: WorldGRID site-view

Another visualization mode is the "site-view" where all resources and monitoring metrics for a site are shown (Figure 2).

There is another plugin at the Nagios server side that is able to generate graphs showing the historical behaviour of several metrics, using the RRD tool.

## 4. CONCLUSIONS

The EDT-monitor tool has been used to spot several problems during the WorldGRID operations, such as malfunctioning Resource Brokers or Information Servers, sites not correctly configured, job dispatching problems, etc. This paper gave an overview of the package, its features and scalability solutions and we reported on the experience acquired and the benefit that a GRID operation center would gain from such a tool.

## References

[1]  iVDGL project
     http://www.ivdgl.org
[2]  DataGrid project
     http://www.edg.org
[3]  CrossGrid project
     http://www.crossgrid.org
[4]  DataTAG project
     http://www.datatag.org
[5]  Nagios monitoring tool
     http://www.nagios.org